\documentclass[twocolumn,prb]{revtex4-2}
\usepackage{titlesec}
\usepackage{subfigure}
\usepackage{amsmath}
\usepackage{amsfonts}
\usepackage{amssymb}
\usepackage{feynmp}
\usepackage{graphicx}
\usepackage{setspace}
\usepackage{comment}
\usepackage{dsfont}
\usepackage{color}
\usepackage[pdftex,colorlinks=true,linkcolor=blue,citecolor=blue]{hyperref}

\begin{document}

\title{Topological states constructed by two different trivial quantum wires}
\author{Jing-Run Lin}
\author{Linxi Lv}
\author{Zheng-Wei Zuo}

\affiliation{School of Physics and Engineering, Henan University of Science and Technology, Luoyang 471023, China}
\date{\today}

\begin{abstract}

The topological states of the two-leg and three-leg ladders formed by two trivial quantum wires with different lattice constants are theoretically investigated. Firstly, we take two trivial quantum wires with a lattice constant ratio of 1:2 as an example. For the symmetric nearest-neighbor intra-chain hopping two-leg ladder, the inversion symmetry protected topological insulator phase with two degenerate topological edge states appears. When the inversion symmetry is broken, the topological insulators with one or two topological edge states of different energies and topological metals with edge states embedded in the bulk states could emerge depending on the filling factor. The topological origin of these topological states in the two-leg ladders is the topological properties of the Chern insulators and Chern metals. According to the arrangement of two trivial quantum wires, we construct two types of three-leg ladders. Each type of the three-leg ladder could be divided into one trivial subspace and one topological nontrivial subspace by unitary transformation. The topological nontrivial subspace corresponds to the effective two-leg ladder model. As the filling factor changes, the system could be in topological insulators or topological metals phases.  When the two-leg ladder is constructed by two trivial quantum wires with a lattice constant ratio of 1:3 and 2:3, the system could also realize rich topological states such as the  topological insulators and topological metals with the topological edge states.  These rich topological states in the two-leg and three-leg ladders could be confirmed by current experimental techniques.
\end{abstract}

\maketitle

\section{Introduction}

Since the discoveries of the Berezinskii-Kosterlitz-Thouless transition\cite{Berezinskii71a, Berezinskii71b, KosterlitzThouless72JPC, KosterlitzThouless73JPC, Kosterlitz74JPC}, and integer and fractional quantum Hall effects\cite{klitzingNewMethodHighAccuracy1980,Tsui82PRL} in the last century, the study of topological quantum matters in condensed matter physics has always been interesting and fruitful\cite{asbothShortCourseTopological2016, bernevigTopologicalInsulatorsTopological2013, MoessnerR21Book, Hasan10RMP, QiXL11RMP, Franz15RMP, WenXG17RMP}. Due to the fantastic properties, the search for novel topological states has extended to the photon and phonon fields\cite{Ozawa19RMP,ZhuWW23RRP}. On the other hand, more and more topological states such as higher-order topological phases\cite{Benalcazar17SCI, Benalcazar17PRB} and non-Hermitian topological phases\cite{BergholtzEJ21RMP} have been uncovered. More recently, according to the square-root operation and tensor product of Hamiltonians, the square-root topological phases\cite{ArkinstallJ17PRB,KremerM20NTC,EzawaM20PRR} and multiplicative topological phases\cite{CookA22CP, PalA24PRB, PalA24PRB2} have been identified.

Due to their relative simplicity, the low-dimensional topological quantum models play an important role in exploration of topological quantum states. The one-dimensional (1D) Su-Schrieffer-Heeger (SSH) chain provides perhaps the pedagogical model supporting nontrivial topological phase with two degenerate zero-energy modes and fractional edge charge\cite{suSolitonsPolyacetylene1979, linRealspaceRepresentationWinding2021, liuTopologicalAndersonInsulators2022, liTopologicalPhasesGeneralized2014,  mondragon-shemTopologicalCriticalityChiralSymmetric2014, songAIIIBDITopological2014, liuTopologicalPhasesNonHermitian2019,ZuoZW22PRA}. The Aubry-André-Harper (AAH) model could be topologically nontrivial and appears different topological edge modes \cite{fraxanetTopologicalPropertiesLongrange2021, ganeshanTopologicalZeroEnergyModes2013, langEdgeStatesTopological2012}. 

On the other hand, using the 1D quantum wire as building block, coupled-wire constructions for multi-leg ladders and two dimensional/three dimensional (2D/3D) quantum wire arrays can emerge the many exotic topological states. For example, two symmetry-preserving stacking of trivial quantum wires can realize the topologically nontrivial phases\cite{choiStackingInducedSymmetryProtectedTopological2023}. The symmetric dimerized two-leg ladders are composed of four different lattice points per unit cell manifest the Floquet topological metal phase with the finite-energy edge states\cite{jangjanFloquetEngineeringTopological2020}. It is also found that either staggered or uniform fluxes threading through the ladder holes may change two-leg ladder topological superconductor system from the BDI class in the Altland–Zirnbauer classification to the D class\cite{wangFluxdrivenQuantumPhase2016}. Based on coupled-wire approach and bosonization technology, the quantum Hall states\cite{Kane02PRL, TeoJCY14PRB, Klinovaja14EPJB}, topological superconductor\cite{SeroussiI14PRB, MongRSK14PRX}, fractional Chern insulators\cite{SagiE14PRB, KlinovajaJ15PRB, SagiE14PRB}, and spin liquids\cite{MengT15PRB2, GorohovskyG15PRB} have been constructed. The 2D phase of shifted charge density waves consisting of an array of weakly coupled 1D charge density waves wires, and the fully gapped bulk charge density waves have topological properties, characterized by a nonzero Chern number\cite{zhangEmergentEdgeModes2023}. 3D systems of coupled quantum wires support fractional topological phases composed of closed loops and open planes of 2D fractional quantum Hall subsystems\cite{mengFractionalTopologicalPhases2015}. In addition, the dissipative two-leg ladder model is studied in the non-Hermitian systems\cite{zhouNonHermitianFloquetPhases2020,wuFluxcontrolledSkinEffect2022,zhangZhangLocalizationDynamicsExceptional2024,liangTopologicalPropertiesNonHermitian2022a}. In a recent study, topological metals are constructed by sliding between trivial spinless quantum wire arrays, which is protected by the inversion symmetry\cite{zuoTopologicalMetalsConstructed2023}.

It is interesting to investigate the possible topological states of the quantum ladders and array formed by different quantum wires. Here, we try to utilize two different trivial quantum wires to construct two-leg and three-leg ladders realizing the topological insulators and topological metals states. The rest of this article is organized as follows. In section \ref{inversion}, we discuss the topological quantum phase transition of the two-leg ladder model formed by two trivial quantum wires with a lattice constant ratio of 1:2 from the trivial insulator to the topological insulator protected by the inversion symmetry and characterized by the topological invariant. In section \ref{asymmetric}, we investigate the topological states for the asymmetric two-leg ladder, which has two different nearest-neighbor inter-chain hoppings. The topological insulators and topological metals can emerge in specific parameter regions.  In section \ref{ThreeLeg}, the topological states for two types of the three-leg ladders are analyzed. The Hilbert spaces for the two three-leg ladders could be divided into two independent subspaces, where one is a trivial subspace and the other is a topological nontrivial subspace. The topological nontrivial subspace corresponds to the Hilbert space of the asymmetric effective two-leg ladder. As the filling factor changes, the topological insulators or topological metals emerge. In section \ref{Extension}, we investigate the topological states for the symmetric two-leg ladders, which are constructed by two different trivial quantum wires with a lattice constant ratio of 1:3 and 2:3. In section \ref{Summary}, the conclusion and outlook are given. 

\section{Two-leg ladder}
\subsection{Inversion symmetric topological insulators}\label{inversion}
\begin{figure}[tbp]
\centering 
\includegraphics[width=0.48\textwidth]{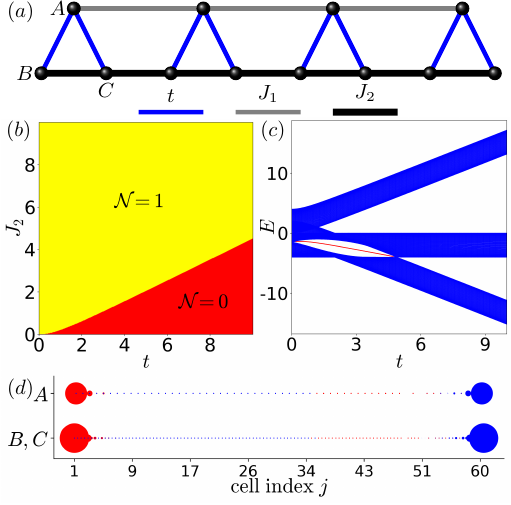}
\caption{(a) The lattice sites of the unit cell for the two-leg ladder are marked as $A$, $B$, and $C$. The inter-chain hopping amplitude is $t$, and the intra-chain hopping amplitudes are $J_1$ and $J_2$. (b) The phase diagram at filling $1/3$ of the two-leg ladder with $J_1 = 1$. The yellow  (red) region represents that the system is in the nontrivial (trivial) topological insulators. (c) Energy spectra as a function of inter-chain hopping amplitude $t$ with system size $L = 60$, $J_1=1$, and $J_2=2$ under OBC. The red dots indicate the topological bound edge states. (d) The density profiles of the two topological bound edge states for filling $1/3$, other parameters are $L=60$, $J_1=1$, $J_2=2$, and $t=2.5$.}
\label{Fig1}
\end{figure}

Firstly, we start with the symmetric intra-chain hopping two-leg ladder (illustrated in Fig.\ref{Fig1}a) of $L$ unit cells, where the amplitudes $t$ of the nearest-neighbor inter-wire coupling  (green lines) between the two different spinless fermion trivial quantum wires are the same. The lattice constant of the upper chain is twice that of the bottom chain. We could write the tight-binding model (set lattice constant of unit cell $a = 1$) in real space as
\begin{eqnarray}\label{eq1}
H_1=\!\!\!\!\!\!&\sum_{j}^{L } (t c^\dagger _{A,j} c_{B,j}+ t c^\dagger _{A,j} c_{C,j}+J_2 c^\dagger _{B,j} c_{C,j})\notag\\
&+\sum_{j}^{L-1} (J_1 c^\dagger _{A,j} c_{A,j+1}+ J_2 c^\dagger _{C,j} c_{B,j+1})+h.c.,
\end{eqnarray}
where $c^\dagger _{\alpha,j} (c_{\alpha,j})$ denotes the creation (annihilation) operator at site $\alpha$ ($\alpha$ stands for the lattice sites $A$, $B$, or $C$ in the unit cell) of the $j$-th unit cell and $t$ is the inter-chain hopping amplitude, whereas $J_1$ and $J_2$ are the intra-chain hopping amplitudes. Without loss of generality, we firstly take all the hopping amplitudes to be real and nonnegative, $J_1,J_2,t\geq 0$.

Assuming periodic boundary condition (PBC) along the length of the two-leg ladder and applying the Fourier transform of the real space creation and annihilation operators, $\psi^\dagger_j=(1/\sqrt{L}) \sum_{k} e^{i k j} \psi^\dagger_{k}$ with $\psi^\dagger_{k}=(c^\dagger _{A} ,c^\dagger_{B},c^\dagger_{C})$, we can express the Hamiltonian in reciprocal space as $H_1=\psi^\dagger_{k} H_1(k) \psi_{k}$ and the reduced bulk Hamiltonian matrix $H_1(k)$ is
\begin{eqnarray}
H_1(k)=\left[
\begin{matrix}
    2J_1 \cos(k) & t &t  \\
    t & 0 &J_2+J_2 e^{ik} \\
    t & J_2+J_2 e^{-ik}  & 0  \\  
\end{matrix}
\right].
\end{eqnarray}

The spectrum of the above Hamiltonian consists of three energy dispersive bands. We can see that  the system does not possess chiral symmetry. Although the chiral symmetry is broken, the inversion symmetry (mirror symmetry, in 1D, the inversion symmetry and mirror symmetry are the same) and time-reversal symmetry are conserved. For this system, the inversion symmetry operator $\mathcal{I}$  is defined by
\begin{eqnarray}
\mathcal{I} =
\left[
\begin{matrix}
    1 & 0 & 0  \\
    0 & 0 & 1   \\
    0 & 1 &0 \\
\end{matrix}
\right].
\end{eqnarray}

The inversion symmetry center lies at the midpoint $A$ between two sites $B$ and $C$ within a unit cell. The inversion symmetry has been widely applied to construct the symmetry-protected topological phases, where $\mathcal{I}H_1(k)\mathcal{I}^{-1} = H_1(-k)$. There are two inversion symmetric invariant momenta $k_{inv} = 0, \pi$ in momentum space. We can define a $\mathbb{Z}$ topological invariant $\mathcal{N}$\cite{hughesInversionsymmetricTopologicalInsulators2011, ChiuCK16RMP}(the Zak/Berry phase can also characterize this topological phase transition):
\begin{eqnarray}
\mathcal{N} = \frac{1}{2} |Tr\{\hat{P}_{k=0}\mathcal{I} \hat{P}_{k=0}^{-1}-\hat{P}_{k=\pi}\mathcal{I} \hat{P}_{k=\pi}^{-1} \}|,
\end{eqnarray}
where $\hat{P}_{k}=\sum_{occ.}|\psi(k) \rangle \langle \psi(k)|$ is the projector operator onto the occupied bands at momentum $k$, $\psi(k) $ is energy eigenstates of $H_1(k)$. The (nonzero) eigenvalues of $\hat{P}_{k_{inv}}\mathcal{I} \hat{P}_{k_{inv}}^{-1}$ is $\zeta(k_{inv})= \pm1$, which is a well-defined parity by eigenstates of $H_1(k_{inv})$. The topological invariant $\mathcal{N}$ represents the absolute value of the difference between the number of negative inversion eigenvalues at $k = 0$ and $\pi$. It generically indicates how many times the insulating gap must close when one takes the atomic limit. According to topological invariant $\mathcal{N}$, we can numerically obtain the phase diagram at the filling $1/3$ of the system, which is shown in Fig.\ref{Fig1}b. The yellow region ($\mathcal{N} = 1$, the Zak phase is $\pi$) represents that the system is in topological insulator phase and the two degenerate edge states would appear in a finite system under open boundary condition (OBC). The red region ($ \mathcal{N} = 0$, the Zak phase is 0) represents that the model is trivial insulator.

Topological invariant is intimately related to the existence of edge states through the bulk-boundary correspondence. A nontrivial topological invariant $\mathcal{N}=1$ implies that a pair of topologically protected degenerate edge states will appear at the boundaries of the system. The energy spectrum of the finite system with OBC is presented in Fig.\ref{Fig1}c, for $J_1=1$ and $J_2=2$. However, due to the chiral symmetry-broken, the energies of the topologically nontrivial edge states are not pinned at zero. The energy spectrum is not symmetric around zero. The red dots indicate the topological edge (localized) states of nontrivial topological insulators. The bulk spectrum is gapped for filling $1/3$. As the hopping amplitude $t$ increases, the system changes from the inversion symmetric topological insulator with nontrivial edge states to trivial insulator. As shown in Fig.\ref{Fig1}d, the two degenerate nontrivial edge states with $L = 60$, $J_1 = 1$, $J_2=2$, and $t=2.5$ are clearly localized in the left and right boundaries. For the filling $2/3$, we can use the similar analysis to obtain the phase diagram and topological states.

\subsection{Topological insulators and topological metals}\label{asymmetric}

\begin{figure}[tbp] 
\centering   
\includegraphics[width=0.48\textwidth]{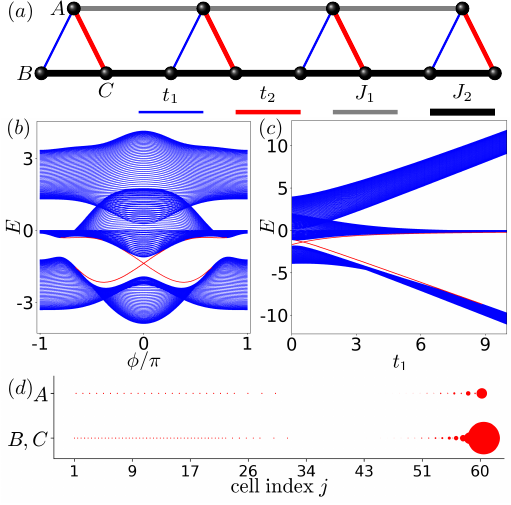}
\caption{(a) The lattice sites of the two-leg ladder are marked as $A$, $B$, and $C$. The inter-wire hopping amplitudes are $t_1$ and $t_2$, the intra-wire hopping amplitudes are $J_1$ and $J_2$.
(b) Energy spectrum of the two-leg ladder model, under OBC as a function of phase factor $\phi$ with system size $L = 60$, and other parameters are $J=1, J_1=1, \lambda=0.94$. The red dots indicate the topological bound end states.  (c) Energy spectrum of the two-leg ladder model under OBC as a function of $t_1$ with system size $L = 60$, and other parameters are $ J_1=1, J_2=1.94, t_2=0.53$.  (d) The spatial distribution of edge state in subgraph (c), the inter-chain hopping $t_1=9$. 
}
\label{Fig2}
\end{figure} 

Now, we investigate the asymmetric inter-chain hopping cases (see Fig.\ref{Fig2}a). The Hamiltonian in real space of the tight-binding model could be written as:
\begin{eqnarray}
H_2=\!\!\!\!\!\!&\sum_{j}^{L } (t_1 c^\dagger _{A,j} c_{B,j}+ t_2 c^\dagger _{A,j} c_{C,j}+J_2 c^\dagger _{B,j} c_{C,j})\notag\\
&+\sum_{j}^{L-1} (J_1 c^\dagger _{A,j} c_{A,j+1}+ J_2 c^\dagger _{C,j} c_{B,j+1})+h.c.,
\end{eqnarray}
where the two different nearest-neighbor inter-chain hoppings are $t_1$ and $t_2$. In momentum space, the Hamiltonian matrix becomes
\begin{eqnarray}
H_2(k)=\left[
\begin{matrix}
      2J_1 \cos(k) &  t_1 &  t_2  \\
      t_1 &  0 & J_2+J_2 e^{ik} \\  
      t_2  & J_2+J_2 e^{-ik}&  0 \\  
\end{matrix}
\right].\label{H2k}
\end{eqnarray}

How to characterize topological properties for the inversion symmetry-broken case is a challenging problem. Here, we view the system as the off-diagonal AAH model ($t_1$ and $t_2$ are the intra-cell hopping and $J_2$ is the inter-cell hopping) with one next-nearest-neighbor hopping $J_2$ (between lattice sites $B$ and $C$) and one third-nearest-neighbor hopping $J_1$ (between lattice sites $A$). Thus, the topological analysis of AAH model can be used to investigate the topological phase for this two-leg asymmetric ladder. As discussed in Refs.\cite{langEdgeStatesTopological2012, alvarezEdgeStatesTrimer2019}, the topological origin of the AAH model is due to the topological properties (non-zero Chern number) for the 2D system. Next, we set the inter-cell hopping amplitudes $t_1 = J [ 1 + \lambda \cos ( 2 \pi / 3 + \phi ) ]$, $t_2 = J [ 1 + \lambda \cos ( 4 \pi / 3 + \phi ) ]$, and $J_2 = J [ 1 + \lambda \cos ( \phi ) ]$. Here, $J$ is the hopping amplitude, and the parameter $\lambda$ is the modulation amplitude of the hopping amplitude. The modulation periodicity of this AAH model is $3$. The conversion of the hopping amplitudes into the modulation parameters $J$ and $\lambda$ are 
\begin{eqnarray}
\left\{
\begin{aligned}
&J = \frac{1}{3} (J_2+t_1+t_2) \\
&\lambda= \pm \frac{2 \sqrt{J_2^2 -J_2 t_1 + t_1^2 -J_2 t_2 - t_1 t_2 + t_2^2}}{\sqrt{J_2^2 +2 J_2 t_1 + t_1^2 +2 J_2 t_2 +2 t_1 t_2 + t_2^2}}  \\
\end{aligned}
\right..
\label{conv}
\end{eqnarray}

\begin{figure}[tbp] 
\centering   
\includegraphics[width=0.48\textwidth]{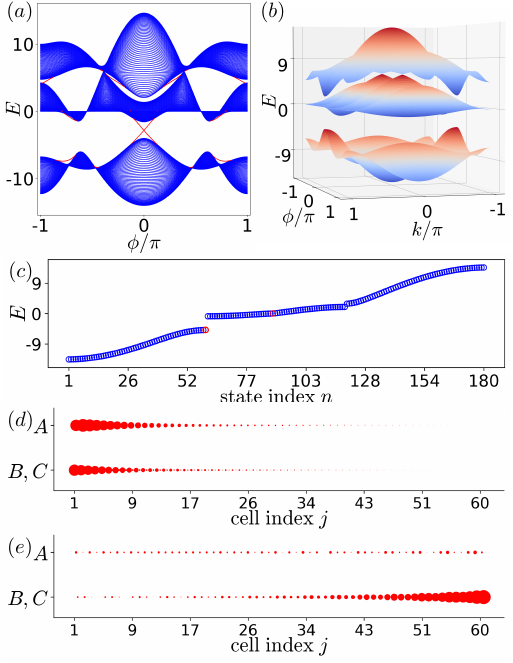}
\caption{(a) Energy spectrum of the two-leg ladder model with OBC as a function of phase factor $\phi$ with system size $L = 60$, and other parameters are $J=1$, $J_1=1$, $\lambda=6$.  The red dots indicate the topological bound end states.  (b) The bulk dispersion relation of panel (a).  (c) Energy spectrum of the two-leg ladder model with OBC with system size $L = 60$.  When phase factor is $\phi=\pi/8$,  other parameters are $ J_1=1, J_2=1+6 \cos(\pi/8), t_1=1+6 \cos(19 \pi /24), t_2=1+6 \cos(35 \pi /24)$.  (d) The spatial distribution of the low-energy edge state of topological insulator at filling $1/3$ for subgraph (c). (e) The spatial distribution of the high-energy edge state of topological metal for subgraph (c).
}
\label{Fig3}
\end{figure} 

Now, assume phase factor $\phi$ as one wave vector, the momentum $k$ and phase factor $\phi$ could be two independent vectors of a 2D system. It means that a family of two-leg ladders, i.e., $\{H_2(k,\phi)|0 < \phi < 2\pi \}$, defines an effective model in two dimensions. Thus, we can numerically calculate the Chern numbers for individual (n-th) bands, which are defined in space ($k, \phi$) over the Brillouin zone $(-\pi\le k < \pi, -\pi \le \phi < \pi )$ as 
\begin{eqnarray}
  {C}_n =\frac{1}{2\pi} \int_{-\pi}^{\pi} \,dk \int_{-\pi}^{\pi} \,d\phi (\partial _{k} A_{\phi}-\partial _{\phi} A_{k}),
\end{eqnarray}
with the Berry connection $A_{\mu}=i \langle \psi(k,\phi) | \partial _{\mu} | \psi(k,\phi) \rangle$ ($ \mu =(k,\phi)$), where $\psi (k,\phi)$ are the Block wavefunctions. 

Due to the next-nearest-neighbor and third-nearest-neighbor couplings, the 2D system can emerge the Chern insulators and Chern metals\cite{cookDoublePerovskiteHeterostructures2014} (indirect nature of bandgap) states (see Fig.\ref{Fig2}b).  On the other hand, as shown in Ref.\cite{alvarezEdgeStatesTrimer2019}, for Chern insulators, there are two or one in-gap edge states in different parameters. For the two-leg asymmetric ladder, as one possible configuration (fixed $\lambda, \phi$, and $J_1$) of the 2D Chern system, the system can emerge the topological insulators and topological metal (topological bound edge states in the continuum\cite{zuoTopologicalMetalsConstructed2023}) states.

In Fig.\ref{Fig2}b, the bulk Bloch states are denoted with blue lines whereas the edge states correspond to red lines. When the inversion symmetry is preserved, these have a crossing at the inversion-symmetry points ($\phi=0$ in Fig.\ref{Fig2}b), indicating the emergence of a pair of degenerate edge states. 
The Chern numbers for the three bands are $(C_1, C_2, C_3)=(1,-1,0)$, we can see the chiral in-gap edge states. To see the evolution behaviors of edge states for the two-leg asymmetric ladder, we plot the energy spectrum as a function inter-chain hopping $t_1$ in Fig.\ref{Fig2}c when the system parameters are $L=60, J_1=1, J_2=1.94, t_2=0.53$. Obviously, when the inversion-symmetry is broken, the two degenerate edge states begin to separate. As the hopping $t_1$ increases, one edge states disappear in the bulk band. The other edge state remains in the energy gap, whose spatial distribution is shown in Fig.\ref{Fig2}d.

By adjusting the parameters ($\lambda, \phi$, and $J_1$), the 2D system can emerge the Chern metals states, which is shown in Fig, \ref{Fig3}a. Because of the indirect nature of band gap, the three energy bands do not touch each other and have decided Chern numbers $(C_1, C_2, C_3)=(1,-2,1)$. In  Fig.\ref{Fig3}b, the bulk dispersion relation of the Chern metals is shown, where the gap for the two top energy bands is indirect. For the two-leg ladder, the system could be in the topological insulators and topological metals phases.  In Fig.\ref{Fig3}c, the energy spectrum of the two-leg ladder under OBC is plotted when the system parameters are $L = 60, J_1=1, J_2=1+6 \cos(\pi/8), t_1=1+6 \cos(19 \pi /24), t_2=1+6 \cos(35 \pi /24) (\phi=\pi/8)$. We can see that there are two edge states with different energies, where one edge state is located at the top of the lowest band, and the other is embedded in the second band. Thus, as the filling factor increases, the system could change from the topological insulator to the topological metal when other parameters are fixed. For the topological edge states in the continuum, they could be identified by the inverse participation ratio (IPR), which is defined by ${\rm IPR}_n =\sum_{j=1}^{3L}  |\psi_j(n)|^4$. The $\rm IPR$ of an extended state scales as $1/3L$, thereby vanishing in the thermodynamic limit, while remaining finite for a localized state. The density profile of the topological edge state at filling $1/3$ is shown in Fig.\ref{Fig3}d and the density profile of topological bound edge state in the continuum is plotted in Fig.\ref{Fig3}e.

\begin{figure}[tbp]
  \centering 
  \includegraphics[width=0.48\textwidth]{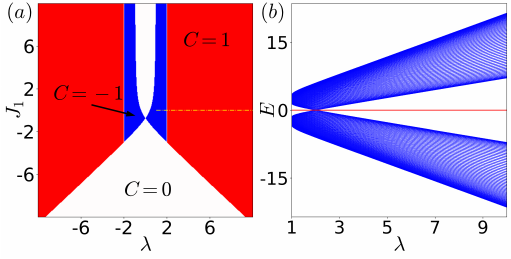}
  \caption{$(a)$ Phase diagram  for the first band in the $\lambda-J_1$ parameter space. The red and blue regions indicate the topological phase. The Chern number is nonzero in the red $(C=1)$ and blue $(C=-1)$ regions;  the Chern number is zero in the white region.  $(b)$ Energy spectra as a function of the modulation parameter $\lambda$ with system size $L = 60$, and other parameters are $J_1 = 0, \phi=\arcsin(1/\lambda)-\pi/6, t_2 =  1 + \lambda \cos ( 4 \pi / 3 + \phi ) , J_2 =  1 + \lambda \cos ( \phi )$ under OBC,  which corresponds to the yellow dashed line in the phase diagram $(a)$.  The red dots in panel $(b)$ indicate the topological bound end states, which coexist with $E=0$ flat band. 
  }
  \label{Fig4}
  \end{figure}
  
Figure \ref{Fig4}a shows the phase diagram for the first band in the $\lambda-J_1$ parameter space. The red region $(C=1)$ and blue region $(C=-1)$  in the phase diagram indicate the topologically nontrivial phase with nonzero Chern number. The topological edge modes exist under OBC in the parameter space of phase factor $\phi$ when the Chern number for the system is nonzero.

On the other hand, the asymmetric two-leg ladder could change into other well-known models by adjusting the parameters $t_1, t_2, J_1,$ and $J_2$. For example, when $t_1 = 0$, $J_1 = 0$, and the modulation parameter $\lambda=\csc (\phi+\pi/6)$ ($|\lambda| \geqslant  1$) in the asymmetric two-leg ladder model, the two-leg ladder reduces to the stub lattice model\cite{molinaFlatBandsPT2015, realFlatbandLightDynamics2017,hudaDesignerFlatBands2020,caceres-aravenaExperimentalObservationEdge2022, pratamaTopologicalEdgeStates2024,pratamaMajoranaRepresentationTopological2024}. The topological properties of the  stub lattice model can be characterized by Chern number. The energy spectrum of the two-leg ladder (stub lattice) under OBC is plotted in Fig.\ref{Fig4}b when the system parameters are 
$L = 60, J_1 = 0, \phi=\arcsin(1/\lambda)-\pi/6, t_2 =  1 + \lambda \cos ( 4 \pi / 3 + \phi ) , J_2 =  1 + \lambda \cos ( \phi ) $, 
which corresponds to the yellow dashed line in the phase diagram Fig.\ref{Fig4}a. The bulk Bloch states are denoted with blue lines, whereas the edge states localized at the boundary correspond to red lines.  For two-leg ladder (stub lattice model), the system obeys the chiral symmetry. Chiral symmetry acts on the Bloch Hamiltonian as $\mathcal{S}H_2(k)\mathcal{S}^{-1} = -H_2(k)$, where $\mathcal{S}$ is the chiral-symmetry operator that  
is defined by
\begin{eqnarray}
\mathcal{S} =
\left[
\begin{matrix}
    -1 & 0 & 0  \\
    0 & -1 & 0   \\
    0 & 0 &1 \\
\end{matrix}
\right].
\end{eqnarray}
In addition to the dispersive bands, the spectra includes a degenerate flat band at zero energy, which coexists with the topological edge states. In the region of $1\leqslant |\lambda|  \leqslant 2 $, the Chern number for the first band is $-1$;  in the region of $ |\lambda| \geqslant  2 $, the Chern number for the first band is $1$. 

In order to decode the topological properties of the two-leg ladder with asymmetric inter-chain hopping, we map it to the general AAH model. In fact, other methods such as graph theory with isospectral reduction approach can also be used to investigate the topological features. Using graph theory, we employ an isospectral reduction approach to transform a Hamiltonian $H$ into an effective Hamiltonian through a dimensional reduction process that preserves the eigenvalue spectrum \cite{bunimovichIsospectralTransformationsNew2014,smithHiddenSymmetriesReal2019,rontgenLatentSymmetryInduced2021,rontgenHiddenSymmetriesAcoustic2023,rontgenTopologicalStatesProtected2024}. Formally, the Hilbert space associated with the Hamiltonian $H$ can be partitioned into a structure set $S$ and its complement $\overline{S}$. The Hamiltonian $H$ takes the form of a block matrix:
\begin{eqnarray}
H=\left[
\begin{matrix}
H_{SS} &  H_{\overline{S} S }   \\
H_{S \overline{S} } & H_{\overline{S}\overline{S}}  \\
\end{matrix}
\right].
\end{eqnarray}
The isospectral reduction $R_S(H, E)$ of $H$ over the set $S$  is then defined through 
\begin{eqnarray}
\mathcal{R}_S(H, E) = H_{SS} + H_{S \overline{S} } (EI - H_{\overline{S}\overline{S}})^{-1} H_{\overline{S} S}
\end{eqnarray}
where $I$ is the identity matrix, and the matrix $H_{\overline{S}\overline{S}} - EI$ is invertible.

Taking the isospectral reduction of $H_2(k)$ over the set $S=\{A,B\}$, we obtain
\begin{eqnarray}
&\!\!\!\mathcal{R}_S(H_2(k), E)\!\!=\!\!\left[
\begin{matrix}
  \frac{t_2^2}{E} + 2 J_1 \cos(k) &  \frac{E t_1 + t_2 J_2}{E}+\frac{t_2 J_2}{E} e^{-ik}  \\[3pt] 
\frac{E t_1 + t_2 J_2}{E}+\frac{t_2 J_2}{E} e^{ik} &\frac{2 J_2^2}{E} (1+\cos(k))  \\
\end{matrix}
\right]\notag\\
&\!\!\!\!\!\!\!\!\!\!\!\!\!\!\!\!\!\!\!\!\!\!\!=s(E) I +
\left[
\begin{matrix}
u(E) &  v(E)+w(E)e^{-ik}   \\
v(E)+w(E)e^{ik} & -u(E)  \\
\end{matrix}
\right],
\end{eqnarray}
where the staggered onsite potential $u(E)=\frac{t_2^2+2J_1E-2 J_2^2(1+\cos(k))}{2E}$, the intracell hopping amplitude $v(E)= \frac{E t_1 + t_2 J_2}{E}$, intercell hopping amplitude $w(E)= \frac{t_2 J_2}{E}$, and $s(E)=\frac{t_2^2+2J_1E+2 J_2^2(1+\cos(k))}{2E}$ all be real. Notably, $\mathcal{R}_S(H_2(k), E)-s(E) I$ takes on the mathematical form of the generalized Rice-Mele(RM) model\cite{riceElementaryExcitationsLinearly1982a}, albeit with energy-dependent hopping amplitudes and energy-dependent onsite potentials. This demonstrates that the isospectral reduction of the asymmetric two-leg ladder reveals a latent RM model. Thus the analysis  methods for the topological properties of the original RM model\cite{kunoNonadiabaticExtensionZak2019,qiTopologicalBeamSplitter2021,haywardEffectDisorderTopological2021,allenNonsymmorphicChiralSymmetry2022a} could be used to the latent RM model and the asymmetric two-leg ladder. Numerical calculations show the two approaches obtain the same phase diagram for the Hamiltonian $H_2$ (Eq. \ref{H2k}) of the asymmetric two-leg ladder.

\section{Three-leg ladder}\label{ThreeLeg}

\begin{figure}[tbp]
\centering
\includegraphics[width=0.48\textwidth]{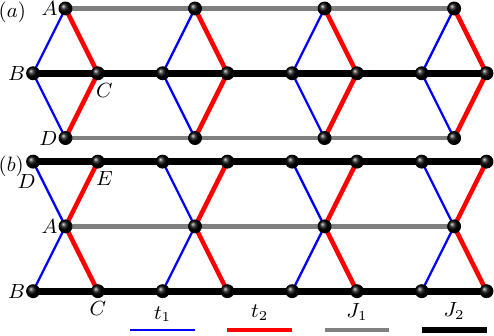}
\caption{(a)The lattice sites of the type I three-leg ladder is marked as $A$, $B$, $C$, and $D$. (b)The lattice sites of the type II three-leg ladder are marked as $A$, $B$, $C$, $D$, and $E$. }
  \label{Fig5}
\end{figure}

\begin{figure}[tbp] 
\centering
\includegraphics[width=0.48\textwidth]{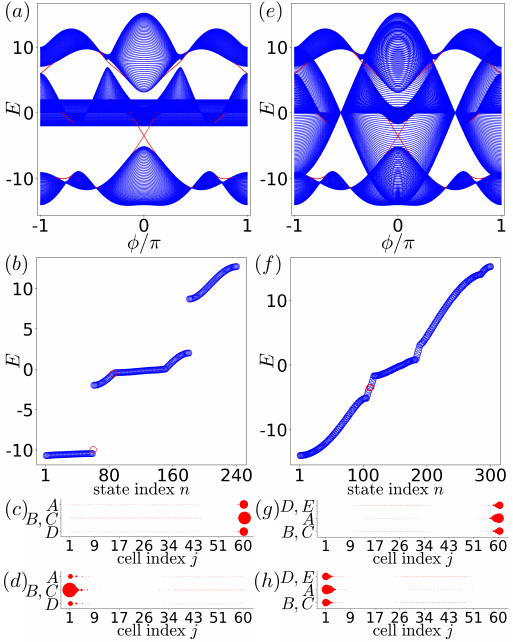}
\caption{(a,e) Energy spectrum of the type I three-leg ladder (a) and the type II three-leg ladder(e) under OBC as a function of phase factor $\phi$ with system size $L = 60$,  and other parameters are $J=1, J_1=1, \lambda=6$. (b) Energy spectrum of the type I three-leg ladder in the real-space under OBC with system size $L = 60$,  when the system parameters are $ J_1=1, J_2=1+6 \cos(3\pi/4), t_1=1+6 \cos(17 \pi /12), t_2=1+6 \cos(25 \pi /12)$, and $\phi=3\pi/4$. (c) The spatial distribution of the low-energy edge state of topological insulator at filling $1/3$ for the subgraph (b). (d) The spatial distribution of the high-energy edge state of topological metal for the subgraph (b). (f) Energy spectrum of the type II three-leg ladder in the real-space under OBC with system size $L = 60$, when the system parameters are $ J_1=1, J_2=7, t_1=1+6 \cos(2 \pi /3), t_2=1+6 \cos(4 \pi /3)$, and $\phi=0$.  (g, h) The spatial distribution of the two degenerate edge states of topological metal for subgraph (f).}
\label{Fig6}
\end{figure} 

Motivated and encouraged by the above results, we investigate the three-leg ladder system. When the two-leg ladder is coupled with the third trivial quantum wire, we can obtain two different types of three-leg ladders.  In the type I three-leg ladder (Fig.\ref{Fig3}a), the lattice constant of center chain is twice that of top chain and bottom chain. In the type II three-leg ladder (Fig.\ref{Fig3}b), the lattice constant of top chain and bottom chain are twice that of center chain. 

The tight-binding Hamiltonian of the type I three-leg ladder in real space can be described by
\begin{eqnarray}
\!\!\!\!\!\!\!\!H_I\!=\!\! &\sum_{j}^{L } (t_1 c^\dagger _{A,j} c_{B,j}+t_1 c^\dagger _{B,j} c_{D,j}+ t_2 c^\dagger _{A,j} c_{C,j}+t_2 c^\dagger _{C,j} c_{D,j}\notag\\
&+J_2 c^\dagger _{B,j} c_{C,j}) + \sum_{j}^{L-1} (J_1 c^\dagger _{A,j} c_{A,j+1}+J_2 c^\dagger _{C,j} c_{B,j+1} \notag\\
&+ J_1 c^\dagger _{D,j} c_{D,j+1}) +h.c.
\end{eqnarray}
where $c^\dagger _{\alpha,j} (c_{\alpha,j})$ denotes the creation (annihilation) operator at site $\alpha$ ($\alpha$ stands for the lattice sites  $A$, $B$, $C$, or $D$.) of the $j$-th unit cell. 

Then, we can express bulk Hamiltonian matrix in momentum space as (base $\psi^\dagger_{k}=(c^\dagger _{k, A} ,c^\dagger_{k, B},c^\dagger_{k, C},c^\dagger_{k, D})$)
\begin{eqnarray}
  \!\!\!\!\!\!\!\!\!\!\!\!\!\!\!\!\!\!\!\!\!\!\!H_I(k)\! =\!\! \left[
\begin{matrix}
2J_1 \cos(k) & t_1 &t_2 & 0  \\
t_1 & 0 &J_2+J_2 e^{ik} & t_1 \\
t_2 & J_2+J_2 e^{-ik}  & 0  & t_2 \\  
0  & t_1 &t_2& 2J_1 \cos(k)  \\
\end{matrix}
\right]\!\!. \ 
\end{eqnarray}
If we use a unitary transformation$H_D(k)=Q^{-1} H_I(k) Q$, the Hamiltonian matrix can be block-diagonalized
\begin{align}
H_D(k) =\left(
\begin{array}[c]{cc}
h_1 & \mathcal{O} \\
\mathcal{O} & h_2
\end{array}
\right),
\end{align}
where $h_1=2J_1 \cos(k)$, the unitary matrix $Q$  and $h_2$ are 
\begin{eqnarray}
&Q=\frac{1}{\sqrt{2}} \left[
\begin{matrix}
-1  & 1 &0  & 0  \\
  0  & 0 &\sqrt{2} & 0  \\
  0  & 0 &0 & \sqrt{2}\\
  1  & 1 &0 & 0  \\
\end{matrix}
\right], \\
&h_2=\left[
\begin{matrix}
      2J_1 \cos(k) &  \sqrt{2} t_1 &  \sqrt{2}t_2  \\
      \sqrt{2} t_1 &  0 & J_2+J_2 e^{ik} \\  
      \sqrt{2} t_2  & J_2+J_2 e^{-ik}&  0 \\  
\end{matrix}
\right].
\end{eqnarray}
Thus, the whole Hilbert space becomes separable. It means that the block diagonalization separates the Hilbert space of the three-leg ladder into two independent subspaces. The $h_1$ is a trivial subspace. The $h_2$ is a topological nontrivial subspace, which is similar to the Hilbert space of the asymmetric two-leg model (Eq.\ref{H2k}) with the effective inter-chain hoppings $\sqrt{2} t_1$ and $\sqrt{2} t_2$. Thus, the analysis method of section \ref{asymmetric} can be applied to this nontrivial subspace.

For the type II three-leg ladder, we can write the tight-binding model as
\begin{eqnarray}
H_{II}\!=\!\!\!\!\!&\sum_{j}^{L-1} (J_1 c^\dagger _{A,j} c_{A,j+1}+ J_2 c^\dagger _{C,j} c_{B,j+1}+ J_2 c^\dagger _{E,j} c_{D,j+1})\notag\\
&+\sum_{j}^{L } (t_1 c^\dagger _{A,j} c_{B,j}+t_1 c^\dagger _{A,j} c_{D,j}+ t_2 c^\dagger _{A,j} c_{C,j}+t_2 c^\dagger _{A,j}c_{E,j}\notag\\
&+J_2 c^\dagger _{B,j} c_{C,j}+J_2 c^\dagger _{D,j} c_{E,j})+h.c. \ .
\end{eqnarray}

The bulk Hamiltonian matrix in momentum space could be written as
\begin{widetext}
\begin{eqnarray}
H_{II}(k)=\left[
\begin{matrix}
2J_1 \cos(k) & t_1 &t_2 & t_1 &t_2   \\
t_1 & 0 &J_2+J_2 e^{ik} & 0 &0  \\
t_2 & J_2+J_2 e^{-ik}  & 0  & 0 &0  \\  
t_1  & 0 &0 & 0 &J_2+J_2 e^{ik} \\
t_2  & 0 &0 & J_2+J_2 e^{-ik} &0  \\
\end{matrix}
\right].
\end{eqnarray}

Using a similar unitary transformation$H_Q=Q_2^{-1} H_{II}(k) Q_2$, we can change the Hamiltonian matrix into block-diagonalized
\begin{eqnarray}
H_Q=\left[
\begin{matrix}
2J_1 \cos(k) & \sqrt{2}t_1 &\sqrt{2}t_2 &0 &0  \\
\sqrt{2}t_1 & 0 &J_2+J_2 e^{ik} & 0 &0  \\
\sqrt{2}t_2 & J_2+J_2 e^{-ik}  & 0  & 0 &0  \\  
0  & 0 &0 & 0 &J_2+J_2 e^{ik} \\
0  & 0 &0 & J_2+J_2 e^{-ik} &0  \\
\end{matrix}
\right]
=\left(
\begin{array}[c]{cc}
h_2 & \mathcal{O} \\
\mathcal{O} & h_3
\end{array}
\right),
\end{eqnarray}
\end{widetext}
where the unitary matrix $Q_2$ and $h_3$ are
\begin{eqnarray}
Q_2=\frac{1}{\sqrt{2}}\left[
\begin{matrix}
\sqrt{2} & 0 &0 & 0 &0  \\
0  & 1 &0 & 1 &0  \\
0  & 0 &1 & 0 &1  \\
0  & 1 &0 & -1 &0  \\
0  & 0 &1 & 0 &-1  \\
\end{matrix}
\right], 
\end{eqnarray}
\begin{eqnarray}
h_3=\left[
  \begin{matrix}
    0 &J_2+J_2 e^{ik} \\
      J_2+J_2 e^{-ik} &0  \\  
  \end{matrix}
  \right].
\end{eqnarray}

So, we can see that $h_3$ is a trivial space. The $h_2$ is a nontrivial subspace, which is the same as that of the type I three-leg ladder. Thus,  there would be similar topological states for the type I and type II three-leg ladders. 

Next, we numerically confirm the existence of topological insulators and topological metals and the topological edge states in the two three-leg ladders. In Fig.\ref{Fig6}a and Fig.\ref{Fig6}e, we plot the energy spectrum of the two three-leg ladders under OBC as a function of phase factor $\phi$, when the system parameters are $L = 60, J=1, J_1=1, \lambda=6$. The Chern numbers of $h_2$ for the three bands are $(C_1, C_2, C_3)=(1,-1,0)$. The nonzero Chern bands will induce different topological states in the two three-leg ladders. For the type I three-leg ladder, we plot the energy spectrum in the real-space under OBC when the system parameters are $L=60, J_1=1, J_2=1+6 \cos(3\pi/4), t_1=1+6 \cos(17 \pi /12), t_2=1+6 \cos(25 \pi /12)$, and $\phi=3\pi/4$. It is easy to see that there are two topological edge states (red circles) with different energies appear. At filling $1/3$, the three-leg ladder is in topological insulator phase with one topological edge state. As the filling factor increases, the system becomes a topological metal state (zero energy around). The spatial distribution of the two topological edge states is shown in Fig.\ref{Fig6}(c,d). For the type II three-leg ladder, the energy spectrum of the type II three-leg ladder in the real-space under OBC is shown in Fig.\ref{Fig6}f, when the system parameters are $L = 60, J_1=1, J_2=7, t_1=1+6 \cos(2 \pi /3), t_2=1+6 \cos(4 \pi /3)$, and $\phi=0$. In this case, inversion symmetry is recovered and the system becomes a topological metal state. At the same time, the two degenerate edge states (red circles) appear, whose density profiles are illustrated in Fig.\ref{Fig6}(g,h).

\section{Two-leg ladder formed by quantum wires with other lattice constant ratios}\label{Extension}

\begin{figure}[tbp]
\centering 
\includegraphics[width=0.48\textwidth]{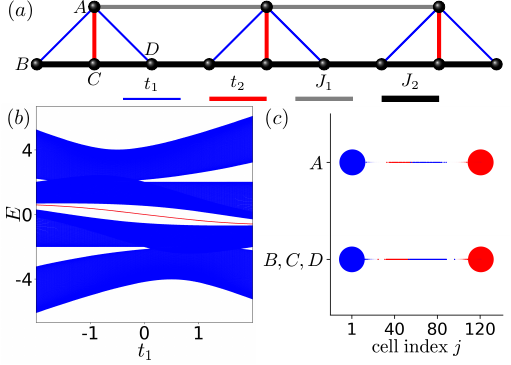}
\caption{(a) The lattice sites of the two-leg ladder are labeled by $A$, $B$, $C$, and $D$. The strength of intra-chain interactions are $J_1$ and $J_2$, and the inter-chain coupling amplitudes are $t_1$ and $t_2$.(b) Energy spectra as a function of inter-chain hopping amplitude $t_1$ with system size $L = 120$, $J_1=t_2=1$, and $J_2=2$ under OBC. The red dots indicate the inversion-symmetry-protected topological edge modes. (c) The density profiles of the two degenerate topological bound edge states for filling $1/2$, other parameters are $L=120$, $J_1=t_2=1$, $J_2=2$, and $t_1=0.5$.}
\label{Fig7}
\end{figure}

Now, we investigate the symmetric intra-chain hopping two-leg ladder formed by two trivial quantum wires with other lattice constant ratios. To begin with, we analyze a two-leg ladder case, where the lattice constant of the upper chain is three times that of the bottom chain, which is illustrated in Fig.\ref{Fig7}a. 
The Hamiltonian of the tight-binding model in real space could be written as:
\begin{eqnarray}
H_a=\!\!\!\!\!\!\!\!\!&\sum_{j}^{L } (t_1 c^\dagger _{A,j} c_{B,j}+ t_2 c^\dagger _{A,j} c_{C,j}+t_1 c^\dagger _{A,j} c_{D,j}\notag\\
&+J_2 c^\dagger _{B,j} c_{C,j}+J_2 c^\dagger _{C,j} c_{D,j})\notag\\
  &+\sum_{j}^{L-1} (J_1 c^\dagger _{A,j} c_{A,j+1}+ J_2 c^\dagger _{D,j} c_{B,j+1})+h.c. \ , 
\end{eqnarray}
where $c^\dagger _{\alpha,j} (c_{\alpha,j})$ denotes the creation (annihilation) operator at site $\alpha$ ($\alpha$ stands for the lattice sites  $A$, $B$, $C$, or $D$.) of the $j$-th unit cell.

Then, we can express bulk Hamiltonian matrix in momentum space as (base $\psi^\dagger_{k}=(c^\dagger _{k, A} ,c^\dagger_{k, B},c^\dagger_{k, C},c^\dagger_{k, D})$):
\begin{eqnarray}
H_a(k)=\left[
  \begin{matrix}
        2J_1 \cos(k) &  t_1 &  t_2 &t_1 \\
        t_1 &  0 & J_2& J_2 e^{ik} \\  
        t_2  & J_2 & 0&  J_2 \\ 
        t_1  & J_2 e^{-ik} & J_2&  0 \\  
  \end{matrix}
  \right].
\end{eqnarray}
For this system, the inversion symmetry operator $\mathcal{I}$ is defined by
\begin{eqnarray}
  \mathcal{I}_a =
  \left[
  \begin{matrix}
      1 & 0 & 0 & 0  \\
      0 & 0 &0 & 1 \\
      0 & 0 & 1 & 0   \\
      0 & 1 &0 & 0 \\
  \end{matrix}
  \right].
\end{eqnarray}

We present the OBC energy spectrum of the finite system in Fig.\ref{Fig7}b, for $J_1=t_2=1$ and $J_2=2$. 
When the filling factor is $1/2$, the nontrivial topological invariant $\mathcal{N}=1$ indicates that a pair of topologically protected degenerate in-gap edge modes will appear at the system boundaries.
The topologically localized boundary modes of nontrivial topological insulators are indicated by red dots. The bulk spectrum is gapped for filling $1/2$. 
As shown in Fig.\ref{Fig7}c, the two degenerate nontrivial edge states with $L = 120$, $J_1 =t_2= 1$, $J_2=2$, and $t=0.5$ are clearly localized in the left and right boundaries.

\begin{figure}[tbp]
\centering 
\includegraphics[width=0.48\textwidth]{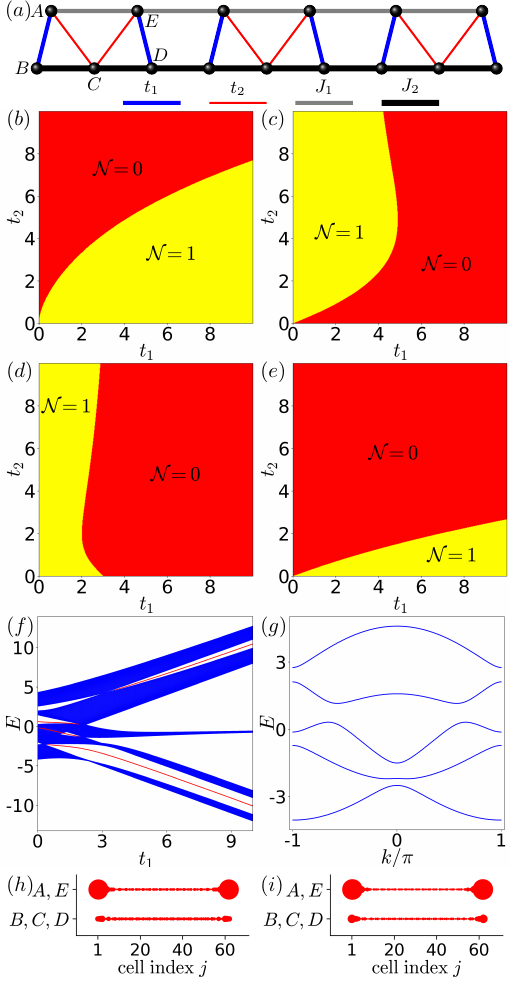}
\caption{(a) The lattice sites of the unit cell for the two-leg ladder are termed $A$, $B$, $C$, $D$, and $E$. The strength of intra-chain interactions are $J_1$ and $J_2$, and the inter-chain coupling amplitudes are $t_1$ and $t_2$. The phase diagram of the two-leg ladder with $J_1=1 $ and $J_2= 2$ at bands as 
(b) lowest energy band, (c) two lower energy bands, (d) three lower energy bands, 
and (e) four lower energy bands. The yellow (red) region indicates that the system is in the nontrivial (trivial) topological insulator phase or topological metal phase.
(f) The  energy spectra under OBC as a function of inter-chain coupling amplitude $t_1$ with system size $L = 60$, $J_1=t_2=1$, and $J_2=2$.
The red dots indicate the topologically protected edge modes. (g) Bloch band structures ($t_1=0.5$) in subgraph (f) indicate the indirect bandgap closure between the second and third bands. 
(h,i) The density profiles of the topological bound edge states of topological metal, other parameters are $L=60$, $J_1=t_2=1$, $J_2=2$, and $t_1=0.5$.}
\label{Fig8}
\end{figure}

Next, we investigate another two-leg ladder system (illustrated in Fig.\ref{Fig8}a) in that the lattice constant of the upper chain is one and a half times that of the bottom chain.
The tight-binding Hamiltonian in real space can be described by
\begin{eqnarray}
&H_b=\sum_{j}^{L } (t_1 c^\dagger _{A,j} c_{B,j}+ t_2 c^\dagger _{A,j} c_{C,j}+t_1 c^\dagger _{D,j} c_{E,j}\notag\\
&+ t_2 c^\dagger _{C,j} c_{E,j}+J_1 c^\dagger _{A,j} c_{E,j}+J_2 c^\dagger _{B,j} c_{C,j}+J_2 c^\dagger _{C,j} c_{D,j})\notag\\
  &+\sum_{j}^{L-1} (J_1 c^\dagger _{E,j} c_{A,j+1}+ J_2 c^\dagger _{D,j} c_{B,j+1})+h.c.
\end{eqnarray}
where $c^\dagger _{\alpha,j} (c_{\alpha,j})$ denotes the creation (annihilation) operator at site $\alpha$ ($\alpha$ stands for the lattice sites  $A$, $B$, $C$, $D$, or $E$.) of the $j$-th unit cell.

Then, the bulk Hamiltonian matrix in momentum space (base $\psi^\dagger_{k}=(c^\dagger _{k, A} ,c^\dagger_{k, B},c^\dagger_{k, C},c^\dagger_{k, D},c^\dagger_{k, E})$) could be written
\begin{eqnarray}
  \!\!\!\!\!\!\!\!\! H_b(k)\!=\!\!\left[
  \begin{matrix}
        0 &  t_1 &  t_2 &0&J_1 +J_1 e^{ik}\\
        t_1 &  0 & J_2&J_2 e^{ik}& 0 \\  
        t_2  & J_2 & 0&  J_2 &t_2\\ 
        0 & J_2 e^{-ik} & J_2 &0&  t_1 \\ 
        J_1 +J_1 e^{-ik}  & 0 & t_2&  t_1&0 \\  
  \end{matrix}
  \right]\!. \ \
\end{eqnarray}

The existence of edge states in this tight-binding model can be explained by nontrivial inversion symmetric topological invariant $\mathcal{N} $. The phase diagram for the invariant $\mathcal{N} $ of the two-leg ladder in the $t_1-t_2$ parameter space for distinct bands: (a) lowest energy band, (b) two lower energy bands, (c) three lower energy bands, and (d) four lower energy bands, which are shown in Fig.\ref{Fig8} b-e, respectively. The yellow region ($\mathcal{N} = 1$, the Zak phase is $\pi$) represents the existence of topological insulator phase or topological metal phase in the system.  The red region ($ \mathcal{N} = 0$, the Zak phase is 0) represents the existence of trivial insulator or metal in the model. In topological insulator phase, a pair of topologically protected degenerate edge states would appear in the finite system under OBC. 
The energy spectrum of the finite system under OBC is presented in Fig.\ref{Fig8}f, for $J_1=t_2=1$ and $J_2=2$.  For small inter-chain hopping $t_1$, the second band and third band overlap, the system becomes inversion symmetry protected topological metal when the Fermi energy fixed between these two bands. In Fig.\ref{Fig8}g, the bulk band structures ($t_1=0.5$) indicate the indirect nature of bandgap closure between the second and third bands.  As shown in Fig.\ref{Fig8}(h,i), the nontrivial edge states of inversion symmetry protected topological metal with $L = 60$, $J_1 =t_2= 1$, $J_2=2$, and $t_1=0.5$ are clearly localized in the boundaries. As the hopping amplitude $t_1$ increases, when the bulk spectrum is gapped for filling factor $2/5$, the system changes from the inversion symmetric topological insulator with nontrivial edge states to trivial insulator.  At fillings $1/5$, and $4/5$  the bulk spectrum are gapped. As the hopping amplitude $t_1$ increases, the system changes from trivial insulator to the inversion symmetric topological insulator with nontrivial edge states. However, for filling $3/5$, the system evolves from  inversion symmetric topological insulator to trivial insulator, as the hopping $t_1$ increases.

When the inversion symmetry is broken in the two two-leg ladders in Figs.\ref{Fig7} and \ref{Fig8}, we can use the same method (mapping the systems to AAH chains) of subsection \ref{asymmetric} to analyze the topological properties. It is straightforward to analyze and inspect the rich topological phases such as topological insulators and topological metals.

\section{Summary and outlook}\label{Summary}

In short, we use the two-leg and three-leg ladders formed by the two different trivial quantum wires to construct the topological states. For the two-leg ladder formed by two different trivial quantum wires with a lattice constant ratio of 1:2, the system can emerge the inversion symmetric topological insulator with two degenerate topological edge states,  the inversion-broken topological insulator with one or two topological edge states of different energies, and topological metals with edge states embedded in the bulk states (topological bound states in the continuum). For the three-leg ladder, using a unitary transformation, we separate the Hilbert space of the system into a trivial subspace and a topological nontrivial subspace. The nontrivial subspace corresponds to the whole Hilbert space of the asymmetric two-leg ladder. The topological insulators and topological metals dependent on the filling factor can also appear in the three-leg ladder.  The constructions for a two-leg ladder using two different trivial quantum wires with a lattice constant ratio of 1:3 and 2:3 could realize the inversion symmetric topological insulator and topological metals with topological edge states. On the other hand, these topological states for the two-leg and three-leg ladders could be realized experimentally by a photonic waveguide array, photonic crystals, topoelectrical circuits, or coupled acoustic resonators. In addition, based on the results of this work, we believe that the constructions for 2D and 3D quantum systems using two or more different trivial quantum wires may realize rich topological states. Furthermore, constructing a two-leg ladder using two trivial quantum wires with a commensurate lattice constant ratio may realize 1D moiré superlattice structures\cite{timmelDiracHarperTheoryOneDimensional2020}. Lastly, using a quasiperiodic incommensurate lattice constant ratio may create 1D bichromatic incommensurate lattices\cite{liMobilityEdgesOnedimensional2017a,vuMoireMottIncommensuration2021}. These constructed structures could emerge elegant physical features.

\emph{Note added:} During the completion stage of this work, we learn that W.C. Shang et al. \cite{shangTopologicalPhasesEdge2024a} discussed the topological phases of the two-chain ladder with different quantum wires, where the topological insulators and topological metals states are also identified.

\emph{Acknowledgements.}--This work was supported by the National Natural Science Foundation of China (Grants No. 12074101, and 11604081). Z.W. Z. is also sponsored by the Natural Science Foundation of Henan (Grant No. 212300410040).

\end{document}